\documentclass[preprint,10pt]{elsarticle}
\usepackage{amsmath}
\usepackage{enumerate}
\usepackage{array}
\usepackage{fullpage}
\usepackage{graphicx}

\journal{Physica A}
\begin{document}
\begin{frontmatter}
\title{Kinetics of node splitting in evolving complex networks}
\author{E.R. Colman}
\ead{Ewan.Colman@brunel.ac.uk}
\author{G.J. Rodgers}
\address{Department of Mathematical Sciences, Brunel University, Uxbridge, Middlesex UB8 3PH, U.K.}
\begin{abstract}
 We introduce a collection of complex networks generated by a combination of preferential attachment and a previously unexamined process of ``splitting'' nodes of degree $k$ into $k$ nodes of degree $1$. Four networks are considered, each evolves at each time step by either preferential attachment, with probability $p$, or splitting with probability $1-p$. Two methods of attachment are considered; first, attachment of an edge between a newly created node and existing node in the network, and secondly by attachment of an edge between two existing nodes. Splitting is also considered in two separate ways; first by selecting each node with equal probability and secondly, selecting the node with probability proportional to its degree. Exact solutions for the degree distributions are found and scale-free structure is exhibited in those networks where the candidates for splitting are chosen with uniform probability, those that are chosen preferentially are distributed with a power law with exponential cut-off.
\end{abstract}

\end{frontmatter}

\section{Introduction}
A considerable literature has been established studying complex networks and their application to various natural and social phenomena. Much of this research concentrates on simple stochastic processes which, when repeated a large number of times, generate complex networks with a variety of topological structures and characteristics. At the simplest level these processes are limited to the addition and removal of nodes and edges, interactions amongst agents in many naturally occurring systems have successfully been described in such a way, consequently a broad range of possible network evolutions are now understood \cite{Albert}.
\newline

In \emph{preferential attachment} a new node is created at each time-step and linked to $m$ existing nodes in the network, by design the likelihood of linking to a node of degree $k$ is proportional to $k$. After many iterations the proportion of nodes which have degree $k$ has been shown to have power law behaviour: $P(k)\sim 2m^{2}k^{-3}$ where $P(k)$ is the proportion of nodes having degree $k$ \cite{scalefree}. Many real-world systems have been explained with this model, most successfully perhaps is the network of citations in scientific publications, papers here are represented by nodes and links are formed between each paper and those papers it cites, empirical data confirms the power law exponent with remarkable accuracy \cite{citations}. An extension of this model incorporates the addition of links between existing nodes \cite{edges}, originally introduced to describe the social network of scientific collaborations though similar variations have also been used to model the interactions of words in human language \cite{dorog}, the degree distribution still follows a power law although it is now composed of two regimes divided by a critical point where the exponent changes.
\newline

In the broader field of statistical mechanics, a substantial body of research concerns the coalescence and fragmentation of clusters of particles, applications in this field span a variety of subjects including astrophysics \cite{astro}, polymerization \cite{polymerization} and aerosols \cite{aerosol}. Despite the diversity of applications the basic model remains the same; two clusters containing either a number, in the discrete case, or mass in the continuous, of identical particles of sizes $x$ and $y$ \emph{coalesce} at a rate $K(x,y)$ into a cluster of size $x+y$. In the discrete setting, the number of clusters of size $x$ at time $t$ denoted by $n(x,t)$ obeys the \emph{Smolochowski coagulation equation},
\begin{equation}
\frac{\partial}{\partial t}n(x,t)=\frac{1}{2}\sum_{y=1}^{x-1}K(y,x-y)n(y,t)n(x-y,t)-n(x,t)\sum_{y=1}^{\infty}K(x,y)n(y,t)
\end{equation}
where the first term on the right hand side accounts for the creation of a cluster of size $x$ from the coalescence of two smaller clusters and the second term accounts for the loss of a cluster of size $x$ when it coalesces with another. Exact general solutions heve not been found, however in the special cases where $K(x,y)=1$ for example, representing two clusters coalescing at each time step, and $K(x,y)=xy$, where clusters coalesce at a rate proportional to their size, exact solutions do exist \cite{coalescence}. Conversely, equivalent equations for fragmentation are constructed in a similar way. If coalescence and fragmentation are simultaneously present in a model then complete fragmentation or complete coalescence into one supercluster can be avoided, in this case the distribution of cluster sizes at large $t$ is independent of $t$. A model of this type has been used to describe the herding behaviour of traders in financial markets \cite{cluster}, here traders are the particles of the system and clusters represent groups of traders sharing information and therefore trading in the same way. The clusters of this model coalesce over time and at random times will rapidly fragment into unclustered individuals. It was shown that the size of the clusters at large $t$ follows a power-law distribution with exponential cut-off, it has also been proposed as a possible reason why variations in share price do not follow a Gaussian distribution \cite{ExactSol}.
The networks presented in this paper extend the lexicon of complex networks by translating the previous model into a network environment. By considering link formation between nodes to be equivalent to coalescence, and by introducing a new process that we shall refer to as ``splitting'' to parallel the fragmentation process described above, we reproduce the cluster size distribution as a network degree distribution.
\newline

\begin{figure}[h]
  \centering
 \includegraphics[width=0.9\textwidth]{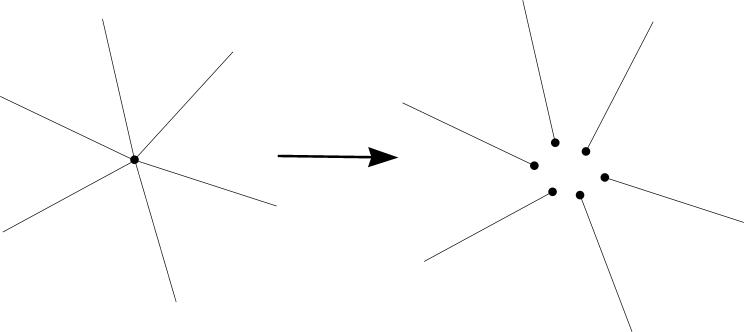}
  \caption{The effect of ``splitting'' a node of degree $6$ in a network.}
  \label{fig:split}
\end{figure}

We define splitting as the replacement of a single node of degree $k$ with $k$ nodes of degree one (see Fig.\ref{fig:split}) and examine the topologies of networks created through this splitting process alongside other growth processes.
The evolution of the networks studied here are driven also by the preferential attachment mechanisms outlined in \cite{prefAtt}, first in Section \ref{nodeSec}, where new nodes are linked to existing nodes in the network chosen with probability proportional to their degree, and secondly in Section \ref{edgeSec} edges are attached between pairs of existing nodes, again with probability proportional to their degree.
\newline

We use $C_{k}(t)$ to denote the number of nodes of degree $k$ at time $t$ and introduce the following two quantities
\begin{equation}
N(t)=\sum_{k=1}^{\infty}C_{k}(t)
\end{equation}
and
\begin{equation}
M(t)=\sum_{k=1}^{\infty}kC_{k}(t)
\end{equation}
where $N(t)$ is the total number of nodes and $M(t)$ is the total degree of the network at time $t$.

\section{Node attachment model}
\label{nodeSec}
At each time step the network may develop in one of the two following ways:
\begin{enumerate}[(a)]
\item With probability $p$ a node is attached by an edge to an existing node, the probability that the end of the edge attaches to a node of degree $k$ is proportional to $k$.

\item With probability $1-p$ a node of degree $k$ is randomly selected and split into $k$ nodes of degree $1$.
\end{enumerate}
At time $t$, for nodes with degree $k\geq2$, $C_{k}(t)$ evolves according to
\begin{equation}
\label{rate1}
\frac{\partial C_{k}}{\partial t}=\frac{p}{M}[-kC_{k}+(k-1)C_{k-1}]-\frac{1-p}{N}C_{k}.
\end{equation}
The first term on the right comes from the loss of a node of degree $k$ that happens when the new edge is attached to it,
the second term comes from the creation of a node of degree $k$ when the new edge attaches to a node of degree $k-1$. The last term
comes from the loss of a node of degree $k$ when it is split into $k$ nodes of degree $1$. For nodes of degree $k=1$,
\begin{equation}
\label{rateC1}
 \frac{\partial C_{1}}{\partial t}=-\frac{p}{M}C_{1}+p+\frac{(1-p)}{N}\sum_{k=2}^{\infty}kC_{k}.
\end{equation}
We can can substitute Eq.(\ref{rate1}) and Eq.(\ref{rateC1}) into the rate equation for $M(t)$,
\begin{equation}
\label{rateM}
 \frac{\partial M}{\partial t}=\frac{\partial C_{1}}{\partial t}+\sum_{k=2}^{\infty}k\frac{\partial C_{k}}{\partial t},
\end{equation}
to verify that $M(t)=2pt$. Similarly, the rate equation for $N(t)$ is found to be
\begin{equation}
\label{nodes}
\frac{\partial N}{\partial t}=(1-p)\left(\frac{M}{N}-1\right)+p.
\end{equation}
Assuming $N(t)$ grows linearly with time i.e $N(t)=\alpha t$, where $\alpha$ is a time independent constant, Eq.(\ref{nodes}) becomes
\begin{equation}
\label{alf}
  \alpha=(1-p)\left(\frac{2p}{\alpha}-1\right)+p
\end{equation}
hence
\begin{equation}
 \alpha=\frac{2p-1+\sqrt{1+4p-4p^{2}}}{2}
\end{equation}
It should be noted that $\alpha$ ranges between $0$ at $p=0$ and increases to $1$ at $p=1$, the slow rate of growth in the size of the network owes to the large likelihood of a node of degree $1$ being selected for splitting resulting in no change to the network. For this reason we find the fastest rate of growth at $p=1$, the usual preferential attachment model without splitting.\newline

The probability $P(k)$ ($\equiv C_{k}/N$) that a randomly selected node will have degree $k$ is solved by substituting $C_{k}(t)=\alpha tP(k)$ into Eq.(\ref{rate1}) giving
\begin{equation}
 P(k)=\frac{1}{2}[-kP(k)+(k-1)P(k-1)]-\frac{1-p}{\alpha}P(k)
\end{equation}
 and so
 \begin{equation}
P(k)=\frac{k-1}{k+2+4(1-p)/(2p-1+\sqrt{1+4p-4p^{2}})}P(k-1)
\end{equation}
thus
\begin{equation}
 P(k)\sim k^{-\gamma}\text{ where } \gamma=3+\frac{4(1-p)}{2p-1+\sqrt{1+4p-4p^{2}}}.
\end{equation}
Comparing this to the case when $p=1$, we see that while the power law structure is not changed by the process of splitting, the exponent can take any value greater than $3$ depending on the value of $p$.

We consider a modification of this model; the same preferential attachment of nodes occurs with probability $p$, however with probability $1-p$ nodes are also preferentially selected for splitting with probability proportional to their degree. The rate equations are
\begin{equation}
\frac{\partial C_{k}}{\partial t}=\frac{p}{M}[-kC_{k}+(k-1)C_{k-1}]-\frac{1-p}{M}kC_{k}
\end{equation}
for $k\geq2$ and
\begin{equation}
 \frac{\partial C_{1}}{\partial t}=-\frac{p}{M}C_{1}+p+\frac{(1-p)}{M}\sum_{k=2}^{\infty}k^{2}C_{k},
\end{equation}
following similar analysis to above the distribution of node degrees in this network is found to follow
\begin{equation}
 P(k)\sim p^{k-1}k^{-\gamma}\text{ where } \gamma=1+2p
\end{equation}
where the power law behaviour is only recovered when $p=1$ and we return to the Barabasi-Albert network described in \cite{prefAtt}.
\subsection{Components}
We compute the number of components $Q(t)$ of the network, that is how many disconnected pieces there are at time $t$, this changes according to the rate equation
\begin{eqnarray}
\frac{\partial Q}{\partial t}&=&(1-p)\sum_{k=1}^{\infty}(k-1)P(k)\\
&=&(1-p)\left(\frac{M}{N}-1\right)
\end{eqnarray}
since the number of components will increase only when a node of degree $k$ is selected for splitting, causing $Q$ to increase by $k-1$ with probability $P(k)$. Solving gives
\begin{eqnarray}
Q&=&(1-p)\left(\frac{2p}{\alpha}-1\right)t\\
&=&\frac{-1+\sqrt{1+4p-4p^{2}}}{2}t
\end{eqnarray}
concluding that the number of components grows at its fastest rate when $p=1/2$, with $Q(t)=(\sqrt{2}-1)t/2$. The model in the following section can be seen as a modification of the node attachment model; with probability $p$ edges are introduced to the network and each end is attached to a node of degree $k$ with probability proportional to $k$, nodes of degree $1$ are no longer introduced to the network and thus are generated only when splitting occurs (with probability $1-p$), we find that a combination of both processes is necessary for the network to grow indefinitely.
\section{Edge attachment model}
\label{edgeSec}
There are two processes that may occur at each time step in the construction of this network
\begin{enumerate}[(a)]
\item With probability $p$ an edge is attached between two existing nodes, the probability that each end of the edge attaches to a node of degree $k$ is proportional to $k$.

\item With probability $1-p$ a node of degree $k$ is randomly selected and split into $k$ nodes of degree $1$.
\end{enumerate}
The rate equation for the behaviour of the number of nodes with degree $k$, $C_{k}(t)$, is
\begin{equation}
\label{rate2}
 \frac{\partial C_{k}}{\partial t}=\frac{2p}{M}[(k-1)C_{k-1}-kC_{k}]-\frac{1-p}{N}C_{k}
\end{equation}

for $k\geq 2$. The first two terms on the right hand side represent the preferential attachment process seen also in Section \ref{nodeSec}, the last term accounts for the loss of a node of degree $k$ by being selected for splitting. When $k=1$ the rate equation is
\begin{equation}
\label{rate2_C1}
 \frac{\partial C_{1}}{\partial t}=-\frac{2p}{M}C_{1}+\frac{1-p}{N}\sum_{k=2}^{\infty}kC_{k}.
\end{equation}
The first term on the right hand side accounts for the loss of a node of degree $1$ that occurs when the new edge is linked to it,
the second term accounts for the increase caused by splitting a node of degree $k$ into $k$ nodes of degree $1$.
As before,we substitute Eq.(\ref{rate2}) and Eq.(\ref{rate2_C1}) into Eq.(\ref{rateM}) to find $M(t)=2pt$, and also
\begin{equation}
 \frac{\partial N}{\partial t}=\frac{\partial C_{1}}{\partial t}+\sum_{k=2}^{\infty}\frac{\partial C_{k}}{\partial t}=(1-p)\frac{M-N}{N}.
\end{equation}
Assuming $N(t)$ grows linearly with time i.e $N(t)=\beta t$, where $\beta$ is a time independent constant, we have
\begin{equation}
 \beta=(1-p)\left(\frac{2p}{\beta}-1\right)
\end{equation}
with the solution
\begin{equation}
 \beta=\frac{p-1+\sqrt{(1-p)(1+7p)}}{2}.
\end{equation}
In contrast to the node attachment model, the limiting values $p=0$ and $p=1$ both produce networks that do not grow with time ($\beta=0$), the rate of increase of $N(t)$ here has its maximum of $\beta=2(2\sqrt{2}-1)/7$ at $p=(3+\sqrt{2})/7$. It is now possible to find the probability $P(k)$ that a randomly selected node will have degree $k$ by solving Eq.(\ref{rate2}). First, note that for large $t$, $C_{k}=\beta tP(k)$, then Eq.(\ref{rate2}) becomes
\begin{equation}
 P(k)=(k-1)P(k-1)-kP(k)-\frac{1-p}{\beta}P(k)
\end{equation}
giving
\begin{equation}
 P(k)=\frac{k-1}{k+1+(1-p)/\beta}P(k-1)
\end{equation}
thus
\begin{equation}
\label{edgedist}
 P(k)\sim k^{-\gamma}\text{ where } \gamma=2+\frac{2(1-p)}{p-1+\sqrt{(1-p)(1+7p)}}.
\end{equation}
Again the edge attachment model can be modified in such a way that the candidates for splitting are selected preferentially, the distribution is found in a similar way to those in the previous sections:
\begin{equation}
 P(k)\sim \left(\frac{2p}{1-p}\right)^{k-1}k^{-\gamma}\text{ where } \gamma=1+\frac{2p}{1-p}.
\end{equation}
Exponential cut-off is present for all values of $p$ except when $p=1/3$, instead the distribution follows a power law with exponent $-2$.
\section{Summary of results}
\begin{table}
\caption{The major results of this paper are reviewed here. ``random splitting'' refers to those models where the candidates for splitting have been selected with equal probability whereas ``preferential'' refers to models that use preferential selection.}
\begin{tabular}{| p{3cm} || c | c | c | }
\hline
 Model & Size (N) & Degree distribution $P(k)$ & Number of components\\
\hline\hline
Node attachment, random splitting & $\left[\frac{2p-1+\sqrt{1+4p-4p^{2}}}{2}\right]t$ &$k^{-\gamma}\text{, } \gamma=3+\frac{4(1-p)}{2p-1+\sqrt{1+4p-4p^{2}}}$&$\frac{-1+\sqrt{1+4p-4p^{2}}}{2}t$\\
\hline
Node attachment, preferential splitting & - &$p^{k-1}k^{-\gamma}\text{, } \gamma=1+2p$&$\frac{-1+\sqrt{1+4p-4p^{2}}}{2}t$\\
\hline
Edge attachment, random splitting & $\left[\frac{p-1+\sqrt{(1-p)(1+7p)}}{2}\right]t$ &$ k^{-\gamma}\text{, } \gamma=2+\frac{2(1-p)}{p-1+\sqrt{(1-p)(1+7p)}}$& -\\
\hline
Edge attachment, preferential splitting & - &$ \left(\frac{2p}{1-p}\right)^{k-1}k^{-\gamma}\text{, } \gamma=1+\frac{2p}{1-p}$& -\\
\hline \hline
\end{tabular}
\label{results}
\end{table}

The degree distributions for each of the models studied here are shown in Table \ref{results} along with the number of nodes and number of components for certain models. We compare the degree distribution of the edge attachment model with preferential selection for splitting with that of the cluster equivalent studied in \cite{ExactSol}. This model evolves by assembling agents into clusters at each time step by the usual coalescence process with probability $a$, or with probability $1-a$ a node is selected and the cluster containing it is split, by which we mean a cluster of size $k$ becomes $k$ clusters of size $1$. It was shown that the distribution of cluster sizes follows a power-law with an exponential cut-off, the power-law exponent is $-5/2$. We observe that the process of attaching edges in the network model is comparable to the coalescence of clusters, and the splitting processes of the two different formulations are similar, the degree distribution also follows a power-law with exponential cut-off and at the value $p=3/7$ the exponent becomes $-5/2$. The models presented in this paper demonstrate splitting as a mechanism to necessitate the creation of nodes, thus the network grows indefinitely contrary to the cluster based model which has a fixed number of agents. Additionally, at each time step the degree of a selected node may only increase by at most one whereas during coalescence a clusters can potentially grow by any amount, these differences prove to be enough not to allow an exact equivalence between the distribution of cluster sizes and the degree distribution.

\section{Conclusion}
Mechanisms that drive the evolution of networks is a topic of interest that is relevant in many different fields, in this work we have introduced one such mechanism and deduced some of the macroscopic qualities of its resulting network. The effect of \emph{splitting} on the growth of a network is not immediately obvious and using two different kinds of preferential attachment models we have deduced some important network properties at large $t$. We have found that a higher frequency of splitting events on the node attachment network speeds the rate of growth of the network while decreasing the proportion of nodes of higher degree. On the edge attachment model splitting drives the creation of new nodes, the maximum growth is reached when $p=(3+\sqrt{2})/7$, as before, the frequency of splitting events decreases the proportion of nodes of high degree.

\section*{Acknowledgements}
This work was supported by EPSRC.

\bibliography{bibfile2}
\bibliographystyle{model1}

\end{document}